\documentclass[fleqn,usenatbib]{mnras}

\usepackage{newtxtext,newtxmath,upgreek}

\usepackage{graphicx}	
\usepackage{subcaption}
\usepackage[T1]{fontenc}
\usepackage{ae,aecompl}
\usepackage{tabularx}

\usepackage{amsmath}	
\usepackage{amssymb}	
\usepackage{natbib, booktabs, multirow, tabularx} 
\usepackage{hhline}

\usepackage{threeparttable}

\usepackage{float}
\restylefloat{figure}

\title[FRB191108]{A bright, high rotation-measure FRB that skewers the M33 halo}

\author[Connor et al.]{%
Liam~Connor$^{1, 2}$,
Joeri~van~Leeuwen$^{2, 1}$,
L.~C.~Oostrum$^{2, 1}$,
E.~Petroff$^{1, 3}$, 
Yogesh~Maan$^{2}$, \newauthor
E.~A.~K.~Adams$^{2, 4}$,  
J.~J.~Attema$^{5}$, 
J.~E.~Bast$^{2}$, 
O.~M.~Boersma$^{1, 2}$, 
H.~D{\'e}nes$^{2}$, \newauthor
D.~W.~Gardenier$^{2, 1}$,  
J.~E.~Hargreaves$^{2}$,
E.~Kooistra$^{2}$,
I.~Pastor-Marazuela$^{1, 2}$,
R.~Schulz$^{2}$,\newauthor
A.~Sclocco$^{4}$,  
R.~Smits$^{2}$,
S.~M.~Straal$^{6, 7}$,
D.~van~der~Schuur$^{2}$, 
Dany~Vohl$^{2}$, \newauthor
B.~Adebahr$^{8}$,
W.~J.~G.~de~Blok$^{2, 9, 4}$,
W.~A.~van~Cappellen$^{2}$, 
A.~H.~W.~M.~Coolen$^{2}$,  \newauthor
S.~Damstra$^{2}$,
G.~N.~J.~van~Diepen$^{2}$,
B.~S.~Frank$^{10, 9}$,
K.~M.~Hess$^{2, 4}$,
B.~Hut$^{2}$,  \newauthor
A.~Kutkin$^{2, 11}$,
G.~Marcel~Loose$^{2}$,
D.~M.~Lucero$^{12}$,
\'A~Mika$^{2}$,
V.~A.~Moss$^{13, 14, 2}$,  \newauthor
H~Mulder$^{2}$,
T.~A.~Oosterloo$^{2, 4}$,
M~Ruiter$^{2, 2}$,
H.~Vedantham$^{2}$, 
N.~J.~Vermaas$^{2}$, \newauthor
S.~J.~Wijnholds$^{2}$,
J.~Ziemke$^{2, 15}$
\\
~\\
$^{1}$ Anton Pannekoek Institute, University of Amsterdam, PO Box 94249, 1090 GE Amsterdam, The Netherlands\\ 
$^{2}$ ASTRON, the Netherlands Institute for Radio Astronomy, Oude Hoogeveensedijk 4, 7991 PD Dwingeloo, The Netherlands\\ 
$^{3}$ Veni Fellow\\ 
$^{4}$ Kapteyn Astronomical Institute, PO Box 800, 9700 AV Groningen, The Netherlands\\ 
$^{5}$ Netherlands eScience Center, Science Park 140, 1098 XG, Amsterdam, The Netherlands\\ 
$^{6}$ NYU Abu Dhabi, PO Box 129188, Abu Dhabi, United Arab Emirates\\ 
$^{7}$ Center for Astro, Particle, and Planetary Physics (CAP$^3$), NYU Abu Dhabi, PO Box 129188, Abu Dhabi, United Arab Emirates\\ 
$^{8}$ Astronomisches Institut der Ruhr-Universit\"at Bochum (AIRUB), Universit\"atsstrasse 150, 44780 Bochum, Germany\\ 
$^{9}$ Dept.\ of Astronomy, Univ.\ of Cape Town, Private Bag X3, Rondebosch 7701, South Africa\\ 
$^{10}$ South African Radio Astronomy Observatory (SARAO), 2 Fir Street, Observatory, 7925, South Africa\\ 
$^{11}$ Astro Space Center of Lebedev Physical Institute, Profsoyuznaya Str. 84/32, 117997 Moscow, Russia\\ 
$^{12}$ Department of Physics, Virginia Polytechnic Institute and State University, 50 West Campus Drive, Blacksburg, VA 24061, USA\\ 
$^{13}$ CSIRO Astronomy and Space Science, Australia Telescope National Facility, PO Box 76, Epping NSW 1710, Australia\\ 
$^{14}$ Sydney Institute for Astronomy, School of Physics, University of Sydney, Sydney, New South Wales 2006, Australia\\ 
$^{15}$ Rijksuniversiteit Groningen Center for Information Technology, P.O. Box 11044, 9700 CA Groningen, the Netherlands
}

\date{Accepted 21 09 2020}

\pubyear{2020}

\begin{document}
\label{firstpage}
\pagerange{\pageref{firstpage}--\pageref{lastpage}}
\maketitle

\begin{abstract}
We report the detection of a bright fast radio burst, FRB\,191108,
  with Apertif on the Westerbork Synthesis Radio Telescope (WSRT).
The interferometer allows us to localise the FRB to a narrow 
$5\arcsec\times7\arcmin$ ellipse by
employing both multibeam information within 
the Apertif phased-array feed (PAF) beam pattern, and across different 
tied-array beams.
The resulting sight line 
passes close to Local Group galaxy M33, with an impact parameter of only 18\,kpc
with respect to the core. 
It also traverses the much larger circumgalactic medium of M31, 
the Andromeda Galaxy. We find that the shared plasma of the Local Group galaxies could contribute $\sim$10\% of its dispersion measure of 588\,pc\,cm$^{-3}$.
FRB\,191108
has a Faraday rotation measure of +474\,$\pm\,3$\,rad\,m$^{-2}$, which is too large to be explained 
by either the Milky Way or the intergalactic medium. Based on the more moderate RMs 
of other extragalactic sources that traverse the halo of M33, we conclude that the dense magnetised plasma resides in the host galaxy. 
The FRB exhibits 
frequency structure on two scales, one that is consistent with 
quenched Galactic scintillation and broader spectral structure with $\Delta\nu\approx40$\,MHz. 
If the latter is due to scattering 
in the shared M33/M31 CGM, our results constrain the Local Group 
plasma environment. We found no accompanying persistent radio sources in the
Apertif imaging survey data.
\end{abstract}

\begin{keywords}
fast radio bursts -- observation -- instrumentation 
\end{keywords}

\section{Introduction}
Fast radio bursts (FRBs) are extragalactic radio pulses, of which approximately 110 
have been discovered to date \citep{lorimer07, petrofffrbcat}. 
They are short duration ($\mu$s--ms), bright (0.01--100\,Jy peak flux density), highly dispersed, and relatively common
\citep[$\sim$\,10$^3$ sky$^{-1}$ day$^{-1}$ above 1\,Jy; ][]{cordes2019, petroff_review}.
The most pressing questions in FRB science fall into two broad categories: What causes these mysterious bursts? And, how can they be put to use? 

In the former class of questions, significant progress has been made in 
the past several years. A subset of FRBs has been found to repeat, 
the first of which was the Arecibo-discovered FRB\,121102 
\citep{spitler2014, spitler2016}. Eighteen repeaters
have been detected with the 
Canadian Hydrogen Intensity Mapping Experiment (CHIME) \citep{chime2019r2, chime19-8repeaters,chime2020-repeaters} as well as one from 
ASKAP \citep{Kumar-2019}. It is still unclear 
if the sources that have not been seen to repeat are of a distinct 
class of once-off events, or if their repetition statistics 
(rate, temporal clustering, luminosity function, etc.) are such that they 
are difficult to detect more than once with 
most telescopes \citep[e.g.][]{Kumar-2019}.
Real-time arcsecond localisation has 
allowed for host galaxy identifications, shedding light 
on the variety of galaxies in which FRBs reside 
\citep{bannister19, ravi-2019-DSA}. Very-long-baseline 
interferometry (VLBI) follow up of repeating 
FRBs has provided milliarcsecond localisation, which has been essential in understanding 
the nearby progenitor environment 
\citep{marcote17, chatterjee17, tendulkar17, bassa17, michilli2018, marcote20}.

In the FRB applications category, the theoretical 
proposals that have been put forward  range from intergalactic 
medium (IGM) and circumgalactic medium (CGM) studies 
\citep{mcquinn2014, prochaska-2019a, vedantham2019}, 
to gravitational lensing \citep{munoz2016, eichler-2017} and cosmology \citep{walters-2018}. 
Recently, progress has been made in 
putting such proposals into practice 
\citep{ravi-2016-scintillation, prochaska-2019b}.

In this paper we report the detection of \mbox{FRB\,191108} 
with the  Apertif Radio Transient System (ARTS) on the Westerbork Synthesis Radio Telescope (WSRT). This source 
has a Faraday Rotation Measure RM=+474\,$\pm\,3$\,rad\,m$^{-2}$, 
which is an order of magnitude larger than the expected 
Galactic and IGM contributions. It also passes through the halo 
of Local Group galaxy M33 (The Triangulum Galaxy) with a best-fit impact parameter 
of just 18\,kpc. The M33 halo is embedded in the much-larger galactic 
halo of M31 (The Andromeda Galaxy), which we expect to also 
impact the  propagation of the pulse. 
In Sect.~\ref{sect-pipeline} we briefly describe the discovery pipeline. We present the burst discovery and localization efforts in Sect.~\ref{sect-results}, and discuss rotation measure and repetition constraints in Sect.~\ref{sect-discussion} and conclude in Sect.~\ref{sect-conclusions}.

\section{ARTS pipeline}
\label{sect-pipeline}
The Apertif Radio Transient System (ARTS) searches for radio pulses
using ten 25-m dishes of the WSRT
equipped with the new Apertif phased array feeds \citep[PAFs;][]{ovc10,2019NatAs...3..188A}.
While a full description of
ARTS is provided in \citet{artsso20}, we highlight a number of relevant features below.

For the real-time FRB search,
we beamform the dipoles in each of the PAFs to produce 40 voltage `compound beams' (CBs)
with 300\,MHz of bandwidth centered on a radio frequency of 1370\,MHz. This 
is done at each dish.
The compound beams
are next further beamformed in firmware across the East-West array to 
create 12 tied-array beams (TABs) per compound beam, out of which 
we generate Stokes I, Q, U, and V data-streams at 81.92\,$\upmu$s and 
195\,kHz time and frequency resolution.
As the fractional 
bandwidth of Apertif is high, $\sim$\,0.2, the TABs 
must be recombined in frequency to produce
`synthesised beams' (SBs).
A synthesised beam points in the same direction across the 
300\,MHz band, which is not true of a TAB. 
An overview of this hierarchical beamforming is provided in \citet{artsso20}.
In total, 71 synthesised beams are formed per compound beam, which span the full primary beam field of view (FoV) 
of $\sim$\,0.23\,deg$^2$. The full 40-compound-beam PAF has a FoV of roughly 9\,deg$^2$.
The total 2840 Stokes I synthesized beams 
are then searched in real time by our single-pulse search software 
{\tt AMBER}\footnote{https://github.com/AA-ALERT/AMBER} \citep{2016A&C....14....1S, sclocco-2020}, which runs 
on a dedicated 40-node graphics processing unit (GPU) computing cluster at the WSRT site. 
Data post-processing is handled by the Data Analysis of Real-time Candidates from the
Apertif Radio Transient System 
(DARC ARTS\footnote{https://github.com/loostrum/darc}; \citealt{darc}) pipeline.
Raw candidates are clustered in  dispersion measure (DM), time, pulse width, and 
beam number; and then  sent to a machine learning 
classifier which assigns a probability of the candidate 
being a true FRB \citep{connor-2018b}. While Stokes I data 
is always written to filterbank files on disk, 
the buffered Stokes Q, U, and V data are only saved if
{\tt AMBER} identifies a candidate with a total duration $<$10 ms, a 
signal-to-noise ratio (S/N hereafter) greater than 10, and a DM more than 20\% larger than the predicted value along the line of sight from the YMW16 electron density model \citep{ymw16-model}.


\begin{figure}
  \centering
    \includegraphics[width=0.48\textwidth, trim=0.25in 0.25in 0.25in 0.25in]{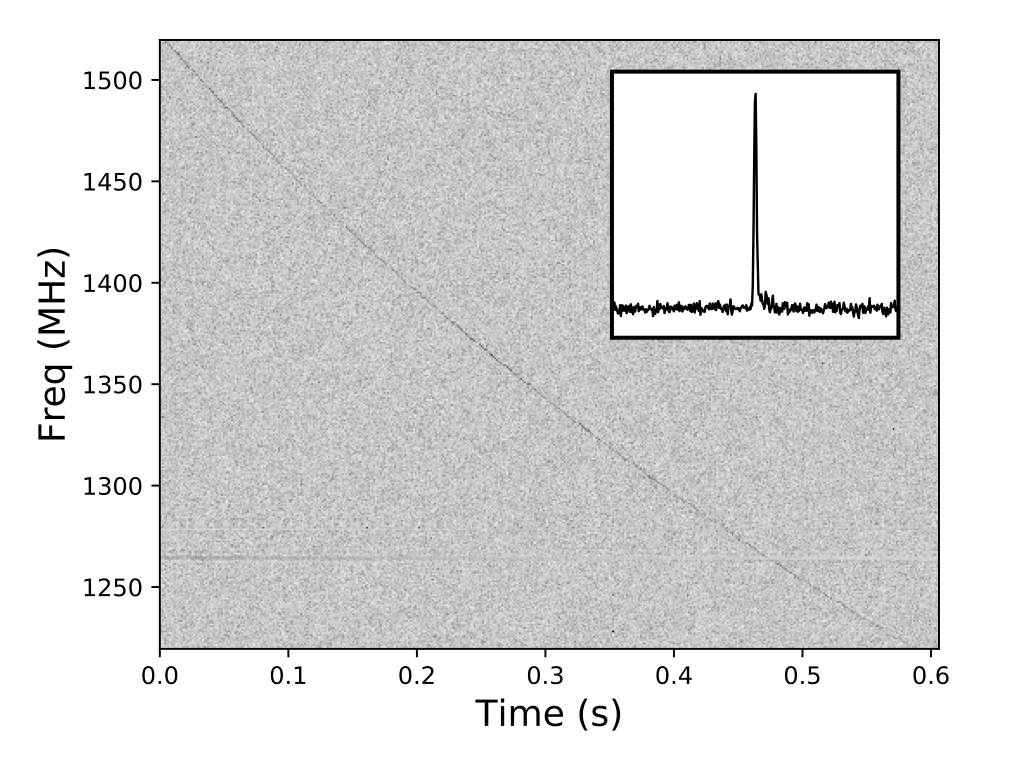}
  \vspace{-2pt}
  \caption{The dispersed dynamic spectrum of \mbox{FRB\,191108} across the ARTS observing bandwidth, and the dedispersed and frequency-averaged pulse profile for 30\, ms of data (inset). The dynamic 
  spectrum has been bandpass corrected and 
  median subtracted, but not RFI cleaned. It 
  is has been binned down to 0.82\,ms time resolution with 0.78\,MHz frequency channels.}
  \label{fig-data}
\end{figure}

\section{Results}
\label{sect-results}

FRB\,191108 was detected in three compound beams, at solar system barycentric UTC 19:48:50.240.
The discovery DM was 588\,pc\,cm$^{-3}$.
Fig.~\ref{fig-data} shows the dynamic spectrum of the dispersed pulse as well as the dedispersed pulse profile. 
The maximum S/N from the real-time 
detection was 60 in compound beam 21 (see Fig.~\ref{fig-localisation}) and our 
machine learning classifier assigned a 
probability of $>$\,99.9$\%$ of it being a real transient \citep{connor-2018b}. 
The {\tt AMBER} detection triggered a dump of the full-Stokes data, allowing us to analyse the polarisation properties of the burst.

\subsection{Polarisation properties}
\label{polarisation}

FRB\,191108 was measured to be roughly 70$\%$ linearly polarised and $\leq$\,10$\%$ 
circularly polarised. It was found to have a rotation measure (RM) of +474$\pm3$\,rad\,m$^{-2}$.
The best-fit RM was obtained by applying a linear least
squares fit to position angle (PA) as a function 
of wavelength squared. The sign was determined by 
verifying that the Crab pulsar had an RM of 
$-$43\,rad\,m$^{-2}$ during an observation the same day. 

Both bandpass calibration and polarisation calibration 
were done using 3C286, a standard calibrator source, which is known 
to have very little circular polarisation. 
We treat the Stokes V value as an upper limit because of uncertainty in the 
polarisation calibration procedure. 3C286 was observed in the same 
compound beam as the FRB, but it was observed in the central TAB, where 
leakage is expected to be minimised. \mbox{FRB\,191108} was found in 
synthesised beam number 37, which is a linear combination 
of non-central TABs. That synthesised beam 
may have slightly different 
leakage properties than the central TAB, which will be better quantified as the system is further calibrated. From the 3C286 on/off observation,  
we solved for a single phase in each down-channelised frequency channel, knowing that the complex $XY$ correlation ought to be purely real if Stokes V is zero. 
We verified that the polarisation calibration solution
agreed with a different method that used the 
FRB itself, which separated the component of 
$\mathbb{I}m\left\{XY\right \}$ that varies with 
$\lambda^2$ from that which does not, since 
Stokes V should not exhibit Faraday rotation under most
circumstances. Fortunately, 
the polarisation rotation does not vary with 
parallactic angle on Westerbork data, as the dishes are on equatorial mounts.
Thus, differences in hour angle between the two observations have no influence.
Still, it is possible that the 
calibration solution is sufficiently different between 
TABs and synthesised beams that the observed 13\,$\%$ 
circular polarisation is spurious. Fortunately, 
Faraday rotation is robust against uncertainty in the 
polarisation calibration solution, because it is difficult 
to mimic a rotation in the Q/U plane that is 
sinusoidal in $\lambda^2$. We are confident in the 
reported value of the rotation measure (RM).

\citet{cho-2020} found that FRB\,181112 exhibited changes 
in its polarisation PA both within and between sub-bursts.
We see no evidence of a swing in the PA across the pulse.
FRB\,121102 was also found to have a flat polarisation PA \citep{michilli2018, Gajjar2018, hessels2019}, as was FRB
180916.J0158+65 \citep[known as R3; ][]{chime19-8repeaters}. This is in contrast to many pulsars 
and it may have interesting implications for FRB emission mechanisms. 
In our case, however, the flat PA may be instrumental. While the 
true PA could be flat across the pulse like previous FRBs, the intrinsic width of \mbox{FRB\,191108} is 
temporally unresolved, meaning any swing in the polarisation PA is 
unobservable; the apparent flat PA across the pulse is the 
time-averaged angle of the true pulse. 
This can lead to depolarisation, because coarse 
temporal sampling and intra-channel dispersion effectively 
add linear-polarisation vectors across the pulse that may point in different directions. 
The depolarisation fraction is

\begin{equation}
  f_{\rm{depol}}\,(\Delta\theta) = 1 - \cos \left (\Delta\theta/2 \right ).
\end{equation}

\noindent Here, $\Delta\theta$ is the PA change across 
the pulse in radians.
Since we observe  $\sim$\,70$\%$ of the FRB emission to be 
linearly polarised, the true pulse 
must be at least as polarised and its 
$\Delta\theta$ cannot be greater than 
$\sim$\,90\,$\degr$. It is possible that \mbox{FRB\,191108} and other 
temporally-smeared FRBs with moderate polarisation fractions have higher intrinsic 
polarisations than inferred.

\begin{figure}
  \centering
    \includegraphics[width=0.48\textwidth, trim=0.0in 0.0in 0.0in 0.0in]{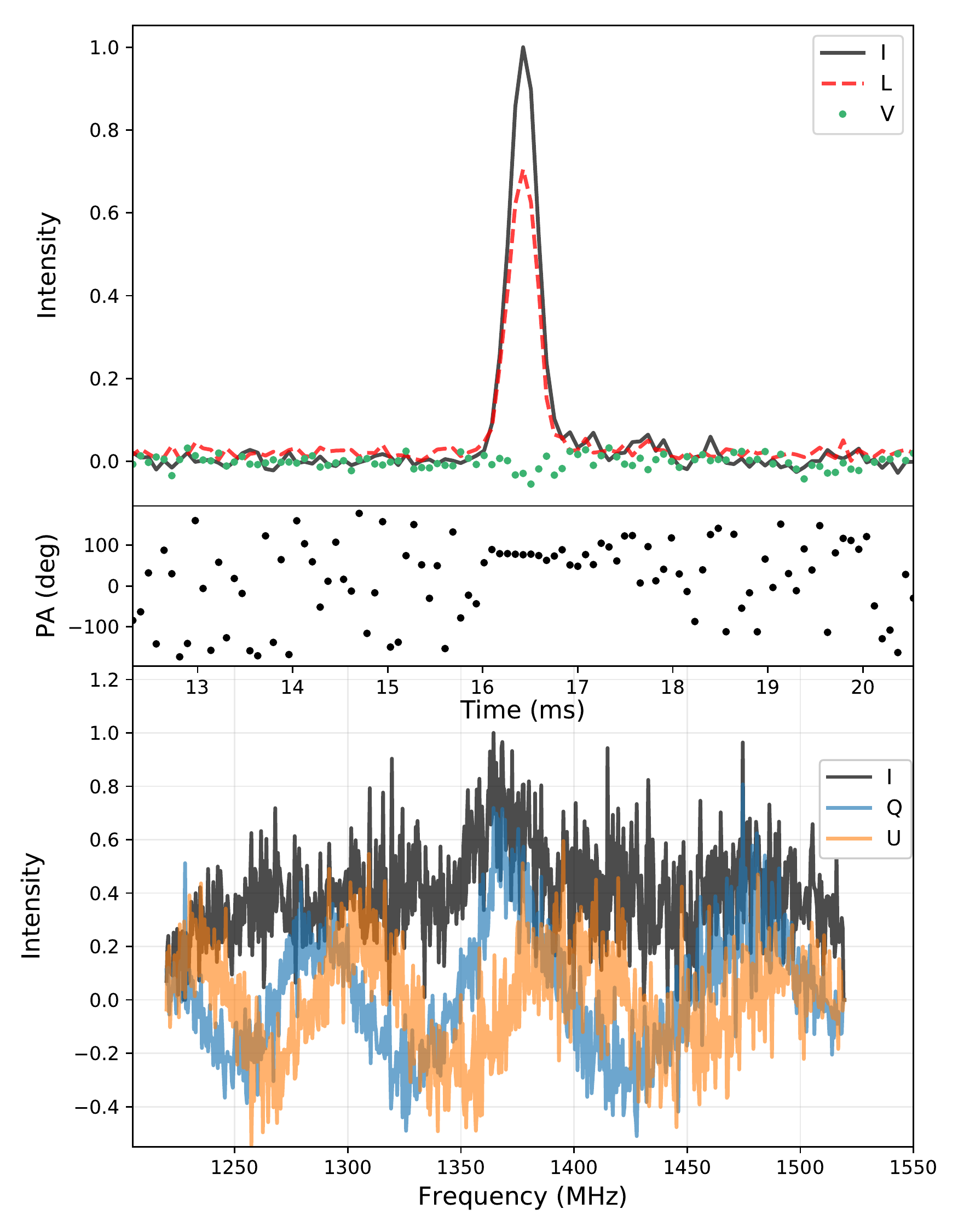}
  \vspace{-2pt}
  \caption{The measured polarisation properties of FRB\,191108. 
   The top panel 
           shows the frequency-averaged pulse profiles after correcting for Faraday rotation in total intensity, 
           I, linear polarisation, L$\equiv\sqrt{\mathrm{Q}^2+\mathrm{U}^2}$, and circular polarisation, V. 
           The middle panel shows a flat PA across the pulse, 
           which could be intrinsic or due to depolarisation, as 
           the true FRB width is temporally unresolved. The bottom 
           panel shows the band-pass corrected frequency spectrum, 
           as well as the Faraday-rotated Stokes Q and U. The best fit 
           RM is +474$\pm3$\,rad\,m$^{-2}$.}
  \label{fig-pol}
\end{figure}

\subsection{Localisation}
\label{sect-localisation}
\begin{figure}
  \centering
    \includegraphics[width=\columnwidth]{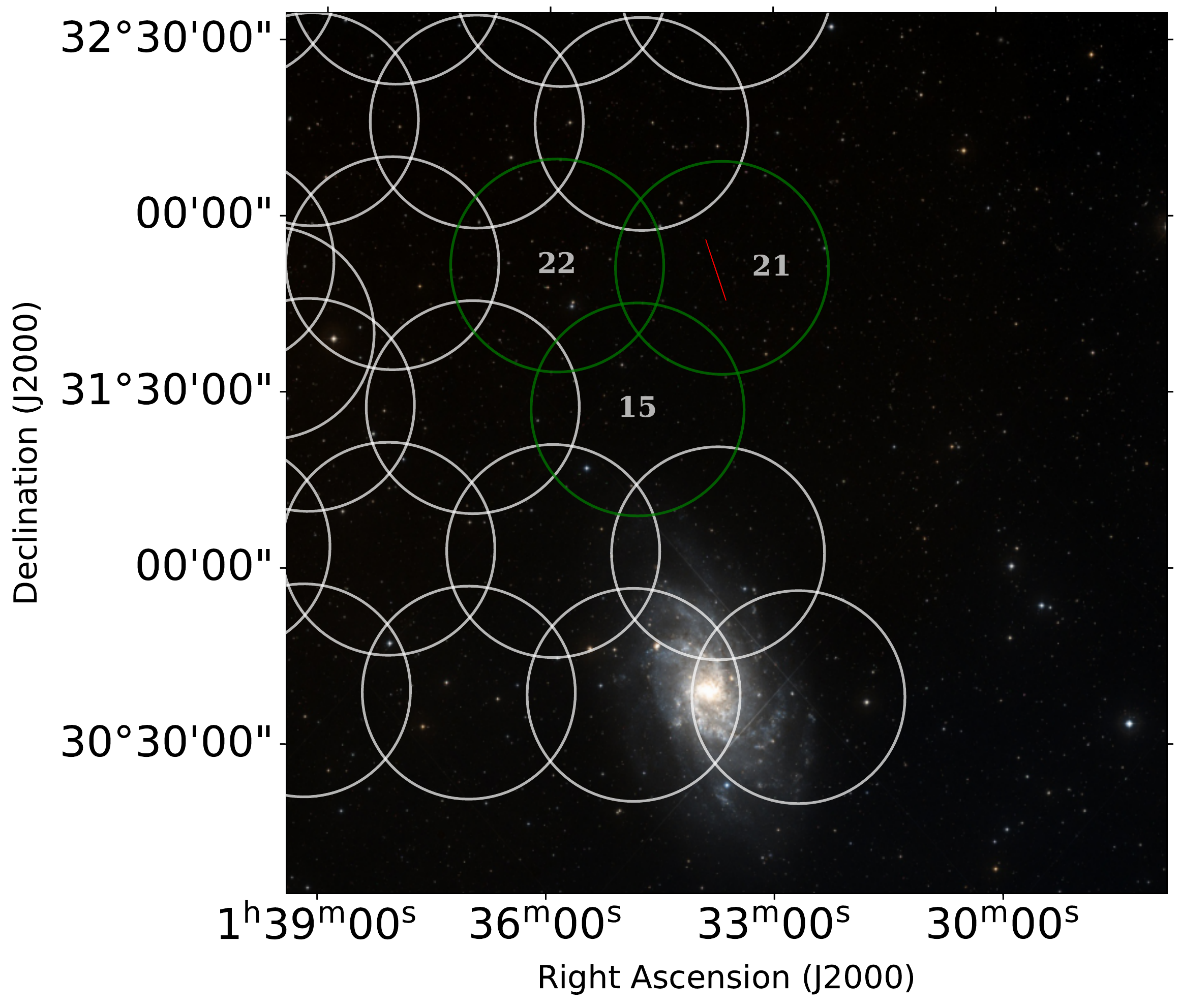}
  \vspace{-2pt}
  \caption{The localisation region of FRB\,191108. The compound beams at 1370\,MHz are shown in
    white (non-detection) and green (detection, with circle opacity in proportion to S/N). The best-fit location is
    shown with a blue cross. The 90\% confidence localisation area is an elongated ellipse, but it is represented in this figure by a red line due to the ellipse's large aspect ratio. The galaxy near the bottom of the figure is M33, which is $1.20\,\pm\,0.05\,\degr$ from the location of the FRB. Background image from the Sloan Digital Sky Survey \citep[SDSS;][]{SDSS}.}
  \label{fig-localisation}
\end{figure}

By combining multibeam information from the 40
overlapping compound beams (CBs) in a PAF, with the interferometric information contained in 
the TABs and synthesised beams (SBs), Apertif can achieve a 
theoretical localisation region of
\begin{equation}
\Omega \approx \frac{30''}{S/N} \times \frac{30'}{S/N}
\end{equation}

\noindent although in practice this will depend on the accuracy of our beam-shape models. 
In order to localise FRB\,191108, we first need to obtain the S/N of the burst in each SB. 
The FRB was initially detected in two neighbouring compound beams,
with the highest S/N in CB 21 (see 
Fig.~\ref{fig-localisation}). Using the 
post-detection optimised DM and timestamp, we measure the S/N of the 
burst in all SBs of CB 21 
and the ones surrounding it. Using a S/N threshold of 8, the FRB was detected 
in CBs 15, 21, and 22, across a total of 48 SBs. 
The highest S/N was 103 in 
SB 37 of CB 21 (hereafter the reference beam).
 
We create a model of the Apertif beam pattern assuming a Gaussian primary beam pattern for each compound 
beam, with a half-power width of $36.3\arcmin$ at 1370\,MHz. Each CB is then scaled using the system-equivalent flux density measured for each CB determined from a drift scan of calibrator source 3C48. Defining a grid of 
$40\arcmin\times40\arcmin$ with a resolution of $1\,\arcsec$ centered on CB 21, 
we generate the TAB response of the 8 equidistant WSRT dishes across this grid and recombine 
these across frequency into 71 SBs per CB. The SBs are 
integrated across frequency, assuming a flat spectral index.
The model is then scaled to the model of the reference beam, resulting in a prediction of 
the S/N ratio between each SB and the reference beam.

Next, we calculate the $\chi^2$ statistic at each grid point. For SBs without a detection, we only include points where
the modelled S/N is above the detection threshold and use the S/N threshold in place of the observed S/N. A 90\%
confidence region is derived from $\Delta\chi^2$ values using the theoretical conversion between confidence level and
$\Delta\chi^2$. The localisation method has been verified using multi-beam detections of giant pulses from the Crab
pulsar and single pulses from PSR J0528+2200, also in CB 21. 

Our method is similar to that employed by CHIME \citep{chime2019r2}. 
The Bayesian approach taken by \citet{bannister2017} for ASKAP is more elaborate, allowing for errors in beam size, sensitivity, and position. However, we note that ASKAP was operating in fly's eye mode. This limits the resolution of a single beam. In contrast, the Apertif data are coherently beamformed across the array, leading to many narrower, higher-resolution beams. This limits the impact of uncertainty in the compound beam positions, as the direction of peak sensitivity of the resulting TABs is dominated by the phase offsets between the dishes. Further improvements to the  determination of Apertif confidence regions, based on several pulsar observations, are in progress \citep{oostrum20}. The localisation code is available online \footnote{\url{https://github.com/loostrum/arts_localisation}}.

The final derived 90\% confidence region is shown in Fig.~\ref{fig-localisation}. The best-fit position (J2000) corresponds to RA=01:33:47, Dec=+31:51:30. The error ellipse has a semi-major axis of $3.5\,\arcmin$ and a semi-minor axis of $2.5\,\arcsec$, with a position angle of $19.5\,\degr$ East of North. The FRB is localised to a region $1.20\pm0.05\,\degr$ from the core of Local Group galaxy M33.
The localisation solid angle of approximately 2100\,square arcseconds (90\,$\%$ confidence) is too large to unambiguously identify a
host galaxy associated with the FRB, even if the 
DM/$z$ relation is to be trusted and utilised \citep{eftekhari-2017}. However, 
as we discuss in Sect.~\ref{sect-repetition}, if \mbox{FRB\,191108} is
found to repeat and is detected at a different parallactic angle, 
we will achieve $\sim$\,arcsecond localisation in both directions 
because the TABs will be at a different position angle on the sky.

\subsubsection{Apertif continuum survey \& radio counterpart}

\begin{figure}[b]
  \centering
    \includegraphics[width=\columnwidth]{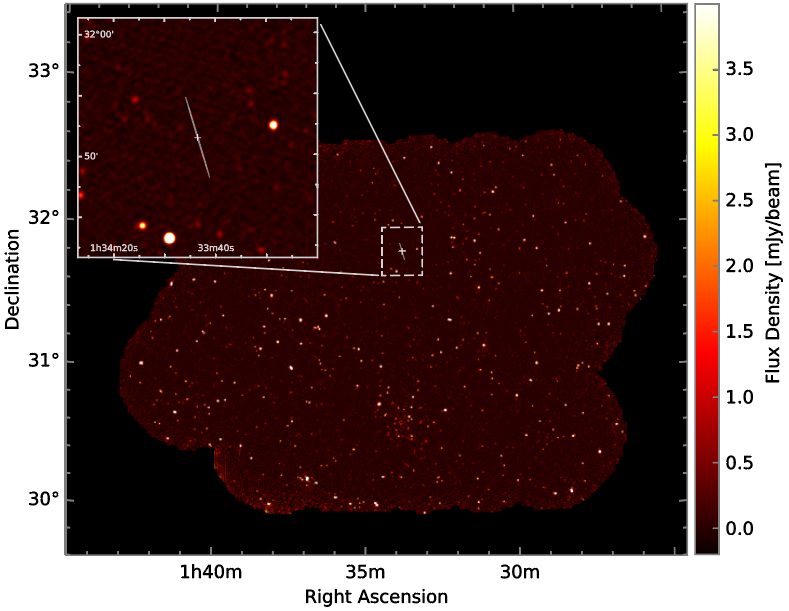}
  \vspace{-2ex}
\caption{A mosaic from the Apertif imaging surveys
          combining 30 compound beams from two adjacent pointings
         around the localisation region.
          The mosaic has a synthesized beam of
$25\mathrm{\,arcsec}\,\times\,25\mathrm{\,arcsec}$.
          In the FRB localisation region, marked by the white ellipse,
no persistent radio counterpart brighter
          than $\sim$350\,$\mu$Jy (5\,$\sigma$ limit) was found.}
  \label{fig-mosaic}
\end{figure}

We have searched for a persistent radio source associated 
with \mbox{FRB\,191108} in continuum images from the Apertif imaging surveys (\citealt{kaa+20}\footnote{\url{https://alta.astron.nl}}).
The mosaic in Fig.~\ref{fig-mosaic} is a combination of 31 compound beams from two survey pointings (191010042 and 191209026) which overlap around the localisation region. The continuum images for the mosaic were made using the top 150\,MHz of the Apertif imaging band (1280--1430\,MHz).
The mosaic covers $\sim$\,9\,deg$^2$ and M33 can be seen in the bottom half of the map. 
We did not find anything within the localisation error region above 5\,$\sigma$ at 71\,$\mu$Jy root mean square
noise.

Radio point sources have a lower on-sky density than 
faint optical galaxies, which decreases the probability of 
chance spatial coincidence and relaxes the localisation requirements for 
radio counterparts \citep{eftekhari-2018}.
The persistent radio source associated with \mbox{FRB~121102}
was roughly 200\,$\mu$Jy 
at $z$\,$\approx$\,$0.2$ at 1\,GHz \citep{chatterjee17}, 
meaning we could have detected  
an equivalent nebula above 3\,$\sigma$ if \mbox{FRB\,191108} were at the same 
distance as FRB\,121102.
This is more nearby than the maximum redshift implied by the extragalactic 
DM of FRB\,191108, which is $z\approx0.52$ (see Sect.~\ref{sect-dm}). Therefore, the host-galaxy ISM or the
dense magnetised plasma contributing to the RM of the FRB
would need to contribute significant DM in order for us 
to detect a persistent source similar to the one associated with 
\mbox{FRB\,121102}. This is not implausible: Using the same 
Galactic halo modelling and DM/$z$ relation employed in this paper, 
the extragalactic DM of FRB\,121102 implies a 
redshift that is 60$\%$ larger than the known value of its host galaxy. 
The Galactic center magnetar, PSR J1745$-$2900, is both 
strongly Faraday rotated (RM$\approx$\,$7\times10^4$\,rad\,m$^{-2}$) and 
dispersed (DM$\approx$1780\,pc\,cm$^{-3}$) near to 
the source, which would make it seem very distant if 
it were bright enough to be seen by an extragalactic observer 
\citep{eatough2013}. Nonetheless, we note that of the five 
unambiguously localised FRBs, no source has a
host-galaxy DM that is known to be significantly more 
than half its extragalactic DM 
\citep{tendulkar17, bannister19, prochaska-2019b, ravi-2019-DSA, marcote20}. 
More host-galaxy localisations will be needed to 
determine how common it is for FRBs to be strongly dispersed 
locally.

If there were a radio source associated with M33 at 840\,kpc, 
we can set an upper limit on its luminosity of 
$\nu L_\nu < 8.5\times10^{31}$\,erg\,s$^{-1}$. At 1400\,MHz, many supernova 
remnants \citep{chomiuk-snr-lumfunction} and HII regions 
\citep{paladini-HII} would have been detectable if they were 
at the same distance as M33. M33 is known to have RGB stars stretching 
$\sim$\,2$\,\degr$ north of the core, nearly three times the radius 
of the classical disk \citep{McConnachie2, McConnachie1}, due to past interactions
with M31. The northern part of M33 also has many HII regions \citep{relano-2013}, but 
most are within 10\,kpc of the core (30 arcminutes below FRB\,191108). 
Therefore, even though it is plausible that 
there would be stellar structure or star formation at the location of 
FRB\,191108, we do not find evidence for a strong Faraday rotating plasma 
associated with M33. These facts, along with the arguments presented in Sect.~\ref{sect-rm-origin}, 
suggest the FRB RM arises in its host galaxy.

\subsection{Time \& frequency structure}
\label{sect-scattering}
We do not find evidence of temporal scattering in 
FRB\,191108 above $\sim$\,80\,$\mu$s. Even though visually there appears to be slightly 
more power after the main peak of the FRB
pulse profile than before it, the detected pulse width is consistent with 
intra-channel dispersion smearing and the sampling time of our instrument. 
We have also fit pulse width as a function 
of frequency and found the data to prefer dispersion 
smearing over scattering.
The latter would result in a 
$\tau\propto\nu^{-4}$ relationship for a single-screen, whereas instrumental 
smearing between channels causes the width 
to scale as $\nu^{-3}$, assuming dispersion smearing is larger 
than sampling time. We find the 
best-fit $\tau(\nu)$ power-law to be $-2.9$, implying that
the pulse is 
temporally unresolved even at 275\,$\mu$s. We also compared our pulse with 
simulation codes {\tt simpulse}\footnote{\url{https://github.com/kmsmith137/simpulse}} 
and {\tt injectfrb}\footnote{\url{https://github.com/liamconnor/injectfrb}}, 
which generate realistically smeared FRBs and account for finite channelisation 
and temporal sampling. We simulated bursts with 
the same DM but varying intrinsic widths, assuming the same 
time and frequency resolution as ARTS, and fit their ``observed'' 
widths with the same pipeline that was used for the FRB. We found that the 
intrinsic width of FRB\,191108, and any scatter-broadening, must be $\lesssim80$\,$\mu$s.

In the top panel of Fig.~\ref{fig-pol}, 
there is excess power after the primary pulse, and 
between 17 and 19 ms the PA appears non-random and consistent with the PA 
of the main pulse. Indeed, when the primary pulse is masked out, we find a 
7.5\,$\sigma$ pulse whose best-fit width is 1\,ms. 
This broader, weaker sub-pulse after the bright, narrow
main pulse has been seen in other FRBs, for example the 
repeating FRB\,180916.J0158+65 (see pulse $d$ in Fig.~1 from \citealt{marcote20})
as well as the first repeater, FRB\,121102 (see pulse $a$ in Fig.~1 from \citealt{michilli2018}).

As argued by \citet{connor-2019}, the observed widths 
of many FRBs are close to the instrumental smearing timescale, 
\mbox{i.e.\ $\sim\sqrt{\tau^2_{\rm DM} + t^2_{\rm samp}}$}, indicating 
that there may exist large numbers of narrow bursts that 
are missed by current search backends. When 
FRBs are coherently dedispersed or observed with high 
time/freq resolution, structure is often revealed on tens of microseconds 
timescales \citep{ravi-2016-scintillation, farah2018, hessels2019}. 
FRB\,191108 may therefore be an example of this population of narrow FRBs
that are often missed without high time and frequency 
resolution backends---something Apertif is fortunate to have.

A least-squares power-law fit was applied 
to the Stokes I frequency spectrum of the FRB, 
yielding a power-law index of $-1.6\pm0.5$. But like 
other fast radio bursts, \mbox{FRB\,191108} is not well described by a 
power law. In the center and top of the band there is a 
factor of $\sim$\,2 of excess power (see the bottom panel of
Fig.~\ref{fig-pol}). Our constraint on the scatter-broadening implies
a lower limit on the Galactic scintillation originated decorrelation bandwidth
to be a few kHz. However, as argued in Section~\ref{sect:scint}, the observed
frequency modulation, with characteristic bandwidth of the order of 40\,MHz,
is unlikely to be due to scintillation. Such bandedness has been seen in
more extreme cases by ASKAP
\citep{shannon2018} and CHIME \citep{chime2019a}, as well as in FRB\,121102
\citep{hessels2019, gourdji2019}. It may prove to be a generic property of
FRB spectra. On the other hand, narrow burst emission only from a few
Galactic neutron stars has been observed to show such bandedness that cannot
be explained by scintillation \citep{Hankins16,Pearlman18,Maan19b}.

\subsection{M33 and M31 halos}
The sky location of \mbox{FRB\,191108} is spatially separated by $1.20\pm0.05\,\degr$ 
and $13.90\pm0.04\,\degr$ from Local Group
galaxies M33 and M31, respectively. 
As M33 is located at a distance of 840\,kpc from the Milky Way,
this translates to an impact parameter of 18\,kpc to the M33 core. 
M31 is approximately 770\,kpc away, meaning \mbox{FRB\,191108} came within roughly 
185\,kpc of Andromeda. Since they are relatively nearby, 
the  circumgalactic medium (CGM) around the two galaxies, as well as the baryonic 
bridge between them, subtend a large angular size. We 
therefore expect the FRB to have traveled through both galaxies' CGM. 
Below we consider how these media might have contributed detectable 
propagation effects to the pulse signature of FRB\,191108. 

\subsubsection{Local Group DM contribution}
\label{sect-dm}
\citet{prochaska-2019a} model the CGM of M31, which is 
large enough to engulf the CGM of M33, as it extends 
$\sim$\,30\,$\degr$. They use a modified Navarro-Frenk-White (NFW) profile and assume 
$M_{\rm halo}^{\rm M31}\approx1.5\times10^{12}$\,M$_\odot$ and 
$M_{\rm halo}^{\rm M33}\approx5\times10^{11}$\,M$_\odot$. The authors also 
consider a  `Local Group Medium (LGM)', 
which models the total intra-group plasma. Using 
Fig.~9 in that paper, \mbox{FRB\,191108} would have an 
additional $\sim$\,40--60\,pc\,cm$^{-3}$ imparted by the halos of 
M33 and M31. 

The hot gas in the Milky Way halo is also expected to 
contribute to the DMs of extragalactic objects. \citet{prochaska-2019a}
estimate a typical contribution of 50--80\,pc\,cm$^{-3}$. 
\citet{yamasaki-2019-halodm} use recent diffuse X-ray observations 
to model the halo DM, and account for the apparent directional 
dependence of emission measure (EM). 
The authors include a hot disk-like halo component 
as well as the standard spherically symmetric halo to 
calculate DM$_{\rm halo}$ as a function of Galactic 
longitude and latitude. 
Using their analytic prescription, we estimate 
the Milky Way halo contribution to be 
$30\pm20$\,pc\,cm$^{-3}$ in the direction 
of FRB\,191108. \citet{keating-2020} find a broader range 
of allowed values for the Galactic halo DM contribution than
previous studies, but also 
favour smaller values. Combining the estimates 
of DM from the Milky Way ISM and halo, along with 
the plasma surrounding M33 and 
M31, the DM of \mbox{FRB\,191108} \textit{beyond} the 
Local Group could be 380--480\,pc\,cm$^{-3}$.  

We use the modelled 
DM/redshift relation from 
\citet{petroff_review}, which is consistent 
with the empirical `Macquart relation' \citep{macquart-2020},

\begin{equation}
    \mathrm{DM_{IGM}}\approx 930\,z\,\,\,\,\mathrm{pc}\,\mathrm{cm}^{-3}
\label{eq-dm-z}
\end{equation}

\noindent and subtracting off the expected Milky Way and Local 
Group DM contribution, the implied redshift upper limit 
on the source is 0.52. If the DM$_{\rm IGM}$/$z$ 
relation is reliable, this is a conservative upper limit 
because it assumes there is zero host-galaxy DM contribution. 
In the case of FRB\,191108, if the Faraday rotation is caused by 
plasma in the host galaxy, there could be non-negligible dispersion 
in the same medium and the true host-galaxy redshift 
would be considerably lower than 0.52.

ASKAP has also found an FRB that appears to pass through 
an intervening halo, coming within 
$\sim$\,30\,kpc of a massive foreground 
galaxy \citep{prochaska-2019b}. This allowed the authors 
to place constraints on the  
net magnetization and turbulence in the foreground galaxy halo,
due to the relatively low RM and dearth of scattering in FRB\,181112.

In our case, the high RM of \mbox{FRB\,191108} does not set a strong 
upper-limit on the halo magnetic field along the line of sight. 
Instead we suggest using the large number of polarised 
extragalactic objects behind M31 and M33 to constrain their CGM 
(see Fig.~\ref{fig-rms}).




\subsubsection{CGM scattering \& scintillation}\label{sect:scint}

We searched for evidence of scattering in both the 
pulse profile and the frequency spectrum of FRB\,191108. 
As shown in Sect.~\ref{sect-scattering}, no temporal 
scattering was found $\gtrsim$\,80\,$\upmu$s. In the frequency 
spectrum, we find structure on two scales: $\sim$\,25\,$\%$
modulations at 40\,MHz and $\sim$\,5\,$\%$ modulations with a 
decorrelation bandwidth of 1--2\,MHz. For comparison, the NE2001 predicts Galactic diffractive scintillation with a correlation bandwidth of 
$\approx$\,1.8\,MHz in the FRB direction \citep{ne2001}.
The 40\,MHz structure could either be intrinsic to the source 
or due to scintillation beyond the Galaxy.

\begin{figure}
  \centering
    \includegraphics[width=0.45\textwidth]{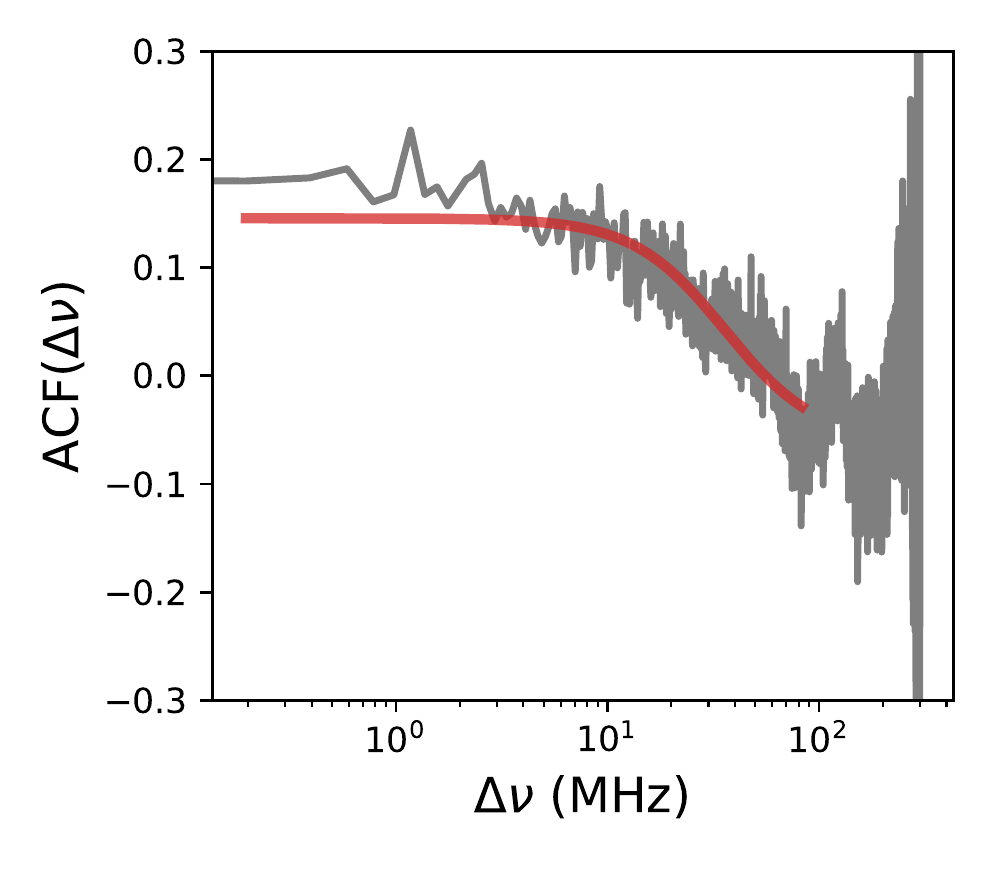}
  \vspace{-2pt}
  \caption{The autocorrelation function 
  of the FRB spectrum, with a best-fit Lorenzian overplotted in red whose de-correlation bandwidth is 40\,MHz.}
  \label{fig-width}
\end{figure}

Our data are sensitive to frequency-domain scintillations with corresponding timescale  in the range 1\,ns\,$\lesssim$\,$\tau$\,$\lesssim$\,500\,ns, 
set by our 300\,MHz band and 0.19\,kHz channel width 
($\tau\approx1/2\pi\Delta\nu$).
To determine the spectral-scale and intensity of scintillations, we compute the autocorrelation function (ACF) of the FRB frequency spectrum 
and fit it with a Lorentzian function \citep{lorimerkramer}, finding a 
de-correlation bandwidth of $\Delta\nu\sim40$\,MHz, 
shown in Fig.~\ref{fig-width}. 
This appears to be dominated by the 
patches of increased brightness around 1370\,MHz and 1500\,MHz, which 
are approximately as wide as the best-fit de-correlation bandwidth. 
This is more than an order of magnitude larger than the expected 
Galactic scintillation bandwidth in the FRB direction.
To search for Galactic scintillation, we
removed the frequency modulation on scales above 20\,MHz 
by subtracting a tenth-order polynomial fit from the data, 
allowing us to look for correlations at smaller $\Delta\nu$. We found significant structure with a correlation scale of a few MHz at the level of 5$\%$ intensity modulation.

For a point-like source, diffractive scintillation leads to 100\,$\%$ modulations 
of the signal, which we do not see in the 1--2\,MHz structure.
The level of modulation would be attenuated if an earlier 
screen has scattered the FRB, leading to angular broadening. 
For a source with 
size $\theta_{\rm sc}$, a scattering screen with a 
diffractive scale $\theta_{\rm d}$ leads to a decrease in the modulation RMS 
of roughly $\,\theta_{\rm d}/\theta_{\rm sc}$. As the intrinsic angular 
size of the FRB is very small, the most natural place for 
this angular broadening is the circumgalactic medium of M33 and/or M31. 
For a characteristic Galactic diffractive scale of $\theta_{\rm d}\approx 0.1\upmu{\rm as}$ \citep{walker-1998}, we should 
expect a $\sim 2\,\upmu{\rm as}$ source size for diffractive scintillation to be attenuated to the 
$\sim$\,5\% levels. 

A $2\,\upmu{\rm as}$ angular broadening at M33 also naturally explains the 40\,MHz-scale structure. A decorrelation bandwidth of 
40\,MHz corresponds to $\tau\approx\,4$\,ns. 
For a scattering screen at $d_{\rm M33}=840$\,kpc, the corresponding angular broadening scale is $\theta_{\rm sc}$\,$\sim2\,\upmu{\rm as}$--- the required value. Therefore, scattering in the halo of M33 parsimoniously accounts for both the suppression of Galactic scintillation as well as the broader 40\,MHz-scale features. It however raises a fresh question as to why the 
the 40\,MHz spectral structure itself, being diffractive in natures, is not observed to be fully modulated. We supply two plausible explanations: (a) We may not be observing a sufficient number of scintels within our bandwidth to measure the total level of modulation. (b) Alternatively, the M33 
scintillation could itself be quenched by a scattering in the 
IGM or CGM of an intervening galaxy. The diffractive scale for the M33 screen in our model is $\theta_{\rm d}\sim 0.03\,\upmu{\rm as}$. Quenching of fully modulated variations by scattering in intervening CGM (unrelated to M31 \& M33) and/or IGM scintillation to the $25\% $ level requires angular broadening at the $0.1\,\upmu{\rm as}$ scale which is within theoretical expectations \citep{koay-macquart-2015,vedantham2019}.

We can derive preliminary constraints on the halo-gas parameters by recognising that the diffractive scale $r_{\rm d} = \lambda/(2\pi\theta_{\rm sc})\sim 10^{11.5}\,{\rm cm}$, denotes the transverse extent over which the rms phase variation is unity. Using \citet[][their equation 4]{coles1987}, for Kolmogovov turbulence, this corresponds to a scattering measure of $\sim 10^{1.3}\,{\rm cm}^{-17/3}$. If the total length through the turbulence is $L\,{\rm cm}$ and the outer scale of turbulence is $L_0\,{\rm cm}$, then using \citet[][their equation 18]{macquart2013}, the dispersion in the electron density is $\left<\delta n_e^2 \right> \sim 10^2L_0^{2/3}/L$. If we further assume that the rms variation in density is equal to the mean density, then we get 
\begin{equation}
    \label{eqn:ne_scat}
    n_e = \left<\delta n_e^2\right>^{1/2}\sim 10^{-2}\left(\frac{L_0}{\rm pc}\right)^{1/3}\left(\frac{L}{\rm pc}\right)^{-1/2}\,{\rm cm}^{-3}.
\end{equation}

The density implied by equation \ref{eqn:ne_scat} is too large to be attributed to the virialized $10^6\,{\rm K}$ circumgalactic gas associated with M33. For example, if we assume that the turbulence is driven by galactic outflows on the scale of $L_0\sim 10\,{\rm kpc}$, then the implied density is $n_e\sim 7\times 10^{-4}(L/100\,{\rm kpc})^{-1/2}$, which is two to three orders of magnitude larger than the expected circumgalactic density of M31 and M33 respectively.

Contrary to simple physical models of virialisation 
in massive dark matter halos, absorption studies 
have found that most quasars that pass within 
$\sim$\,150\,kpc of a foreground galaxy indicate 
the existence of cool ($10^4$\,K) gas 
embedded in a hot ($10^6$\,K) CGM \citep{prochaska14}. It has been argued that 
gas in these environments is prone to fragmentation, leading to a `cloudlet' model 
of the CGM in which sub-parsec cold gas clumps are distributed throughout the 
hot background medium \citep{mccourt17}.
Following the suggestion of \citet{vedantham2019}, we next consider scattering in such cool clumps in the CGM of M31. If the clumps form from cooling instabilities as suggested by \citet{mccourt17}, they will have a density of $\sim$\,$ 10^{-3}\,{\rm cm}^{-3}$, and a length-scale of about 20\,pc which we take to be the outer scale of turbulence. The path length through the cool clumps is given by the virial radius ($200\,{\rm kpc}$ for M31) times the volume fraction, $f_V$. The density constrain from equation \ref{eqn:ne_scat} for this scenario is $n_e\sim 10^{-2.5}(f_V/10^{-3})^{-1/2}\,{\rm cm}^{-3}$, which is comparable to the anticipated value.

In summary, the scattering constrain is roughly consistent with expectations from tiny $10^4\,{\rm K}$ clumps formed via cooling instabilities in the circumgalactic medium of M31. We note two caveats here, however. The sight line to the FRB passes close to a neutral gas bridge connecting M31 and M33 \citep{McConnachie1}. As such, it is unclear if the scattering constraint is probing the specific conditions in neutral bridges or in the general circumgalactic medium. Secondly, 
with such few scintles we cannot know with certainty if the 
$\sim$\,40\,MHz decorrelation bandwidth corresponds to 
scattering or if it is intrinsic to the burst. Future observations
of FRBs intersecting the Local Group medium will 
help answer these questions, including from CHIME where the 
decorrelation bandwidth would be a $\sim$\,30 times narrower.

\begin{table}\centering
\caption{FRB\,191108 parameters. $\dagger$ are values that have been optimised post real-time detection, chosen to maximise S/N in the 
the case of width and DM. 
Our localisation region is an ellipse whose 
semi-major and semi-minor axes do not correspond to RA and Dec, so we 
do not quote uncertainty on those values. For the true sky localisation 
parameterisation, see Sect.~\ref{sect-localisation}. The width listed here is dominated by intra-channel dispersion smearing, but we set an 
upper-limit on its intrinsic width at 80\,$\mu$s.}
\begin{tabular}{lccccc}
                 &                                                    &  &  &  &  \\ \cline{1-4}
Date             & 2019 November 8                                    &  &  &  &  \\
UTC$^{\rm (a)}$               & 19:48:50.471        &  &  &  &  \\
MJD$^{\rm (b)}$        & 58795.830818389                             &  &  &  &  \\
RA (J2000)       & 01h33m47s                                     &  &  &  &  \\
Dec (J2000)      & +31d51m30s                                    &  &  &  &  \\
\cline{1-4}
\\
DM$^\dagger$               & 588.1$\pm$0.1\,pc\,cm$^{-3}$ &  &  &  &  \\
RM               & +474$\pm$3\,rad\,m$^{-2}$ &  &  &  &  \\
Width$^\dagger$ (1370 MHz) & 340$\pm20$\,$\mu$s                         &  &  &  &  \\
Flux density     & 27\,Jy                              &  &  &  &  \\
Fluence    & 11\,Jy\,ms                              &  &  &  &  \\
S/N$_{\rm det}$  & 60                                                 &  &  &  & 
\\
S/N$_{\rm opt}^\dagger$  & 103                                                 &  &  &  & 
\\
\cline{1-4}\\
DM$_{\rm MW}$ (YMW16/NE2001)              & 43 / 52\,pc\,cm$^{-3}$ &  &  &  &  \\
RM$_{\rm MW}$                           & $-$50\,rad\,m$^{-2}$ &  &  &  &  \\
$z_{\rm max}$                           & 0.52 &  &  &  &  \\

\cline{1-4}
\\
\end{tabular}
\begin{tablenotes}
\item $^{\rm (a)}$ At 1370\,MHz.
\item $^{\rm (b)}$ At the solar system barycenter after removal of the DM delay.
\end{tablenotes}
\end{table}

\section{Discussion}\label{sect-discussion}

\subsection{Rotation measure origin}
\label{sect-rm-origin}
\begin{figure}
  \centering
    \includegraphics[width=0.48\textwidth, trim=0.25in 0.25in 0.25in 0.25in]{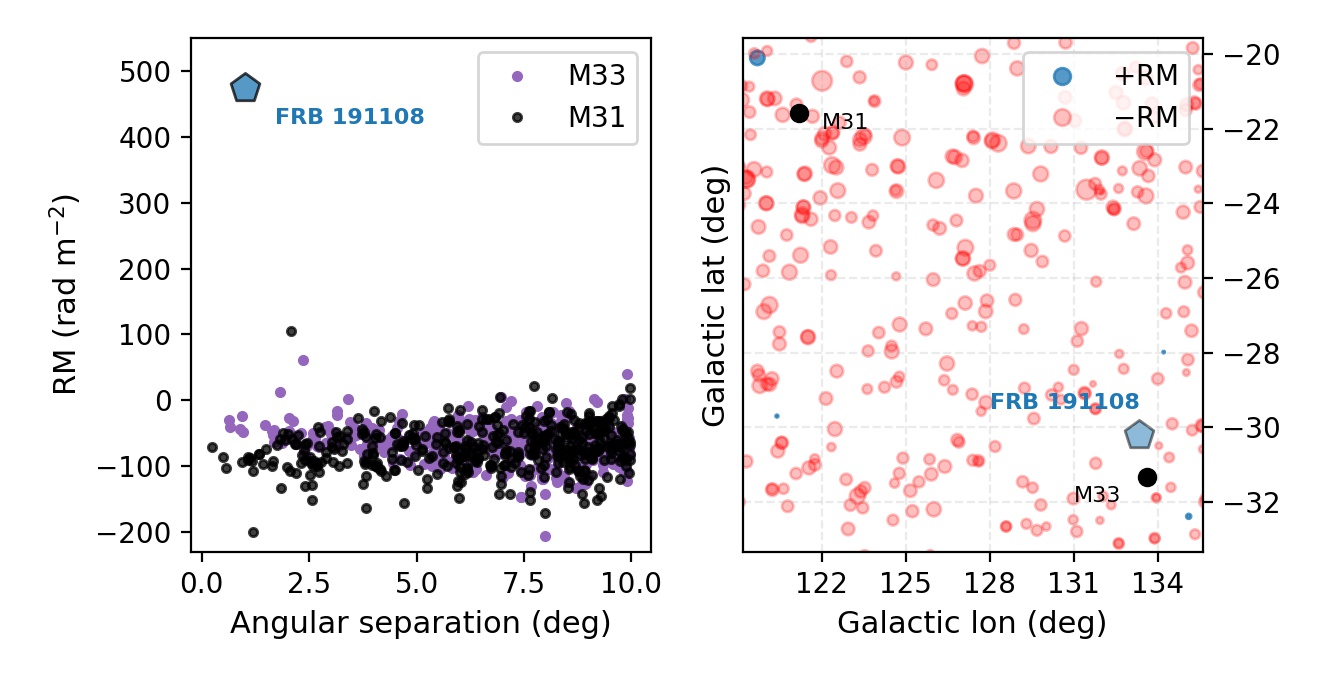}
  \vspace{-2pt}
  \caption{The RMs of extragalactic sources in the 
  direction of the Local Group galaxies M33 and M31. 
  The left panel shows RM vs. angular separation for 
  both M33 (purple) and M31 (black), as well as the FRB which 
  is an outlier both in amplitude and sign. The right panel 
  shows extragalactic sources, where the 
  area of the marker is proportional to |RM| and the colour encodes its sign.
  For size reference, the FRB |RM|=+474\,rad\,m$^{-2}.$}
  \label{fig-rms}
\end{figure}

The observed RM of an FRB can be broken down into several components 
between the observer and source,
\begin{equation}
    \rm{RM}_{\rm obs} = \rm{RM}_{\rm MW}+\rm{RM}_{\rm IGM}+\rm{RM}_{\rm host},
\end{equation}

\noindent where RM$_{\rm MW}$ is the foreground RM from the Galaxy, RM$_{\rm IGM}$ is from the intergalactic medium, and
RM$_{\rm host}$ comes from the host galaxy ISM and the region near the FRB progenitor.
In the case of FRB\,191108, we might also include RM$_{\rm LG}$, the contribution from the Local Group.
This is the contribution of the galactic halos of M33 (Triangulum) and M31 (Andromeda), 
and the broader shared plasma linking the two nearby galaxies with the Milky Way.
The expected 
Milky Way foreground is
$\rm{RM}_{\rm MW}\approx-50$\,rad\,m$^{-2}$
\citep{oppermann2015}. Fig.~\ref{fig-rms} provides an idea of the spatial scatter of this value.
Our observed RM$_{\rm obs}$=$+474\pm3$\,rad\,m$^{-2}$ thus translates to an
estimated extragalactic contribution of approximately 
525\,rad\,m$^{-2}$, which could be up to a couple of times  
larger in the host-galaxy frame due to cosmological redshift.

Such a large extragalactic RM is not expected from the IGM, as it would 
require ordered $\mu$G magnetic fields over gigaparsec scales 
to achieve $10^{2-3}$\,rad\,m$^{-2}$ for typical FRB redshifts. 
No intergalactic magnetic fields have been detected, but they 
are expected to be roughly nG in strength \citep{michilli2018}.

We consider the possibility that the ionised material surrounding M33/M31
could contribute all the required magnetised plasma to account for the RM of the FRB, 
but do not find this compelling for the following reason.
By taking the catalogue of 41632 extragalactic RMs from \citet{oppermann2012}, 
we identify 93 objects that pass within 5\,$\degr$ of 
M33, roughly the angular radius of the 
expected 75\,kpc halo. 93$\%$ of these sources have RMs between $-15$ and $-90$\,rad\,m$^{-2}$---probably dominated by the Milky Way foreground like most 
polarised extragalactic sources---and 
none is larger in magnitude than 100\,rad\,m$^{-2}$. 
In Fig.~\ref{fig-rms} we plot the distribution of extragalactic RMs 
near the Local Group on the sky to demonstrate the extent to which 
FRB\,191108 is an outlier. 

We have also looked at polarised extragalactic 
sources closer to the FRB sight line in the Apertif 
imaging data, which has more sources per solid angle than 
the NVSS RM catalogue \citep{hess2020}. 
We find a picture consistent with the 
Oppermann map \citep{oppermann2015}, in that the distribution 
of RMs clusters around $-50$\,rad\,m$^{-2}$ and no 
point source has an RM as large at FRB\,191108. One source 
is within $\sim$\,30 arcminutes of the FRB's best-fit position 
and likely also intersects the material bridge connecting M33 and M31. Its 
RM is $-72$\,rad\,m$^{-2}$, consistent with the values of 
extragalactic RMs in the surrounding $\sim\,10^\circ\times10^\circ$.

Therefore, unless 
the FRB has a very unusual sight-line and travels through 
a dense magnetoionic region in the M33/M31 halo with the opposite 
magnetic field sign, the absence of strong Faraday rotation in 
other extragalactic polarised sources behind M33
suggests the FRB RM is imparted elsewhere. The dataset plotted 
in Fig.~\ref{fig-rms} and the polarised Apertif imaging data
could still be a useful probe of CGM 
magnetic fields in its own right: the black points in the left 
panel that have a low impact parameter with M31 
show a small gradient such that their amplitude increases towards 
smaller angular separations. Whether this is due to structure in the Galactic 
foreground Faraday field or in the M31 halo could be teased out with a Galactic DM map and we leave it to future work to 
disentangle these effects.

Given we do not expect the large RM of the FRB to be dominated by the 
Milky Way, M33, or the IGM, it is likely that the magnetised plasma 
is in the host galaxy. Using the 
estimated maximum redshift implied by the extragalactic DM, of $z\approx0.52$, and noting that the local RM will be a 
factor of $(1+z)^2$ larger than the observed RM due to cosmological 
redshift, RM$_{\rm host}$ could be of order 10$^3$\,rad\,m$^{-2}$.
Even if the host galaxy contributes significantly to the extragalactic  
DM and the FRB is much closer than the redshift implied by 
Eq.~\ref{eq-dm-z}, the RM would still be much larger than that expected from the ISM of a Milky Way-like galaxy,
unless observed very close to edge-on. 

FRBs are now known to be located in a range of environments spanning 
different galaxy types. 
While there exist examples of polarised FRBs 
without significant Faraday Rotation 
\citep{ravi-2016-scintillation, petroff2017polarized},
now including a repeater \citep{chime2020-repeaters}, several sources appear to 
pass through regions of highly-magnetised plasma, which may be directly 
linked to the FRB progenitor itself (e.g. young supernova remnant). Alternatively, 
FRBs may just be preferentially born in environments that have an abundance 
of sightlines that intersect, say, HII regions.
The first was FRB\,110523, which was detected with the Green Bank Telescope. It had an 
RM of $-186$\,rad\,m$^{-2}$. Like the Apertif-discovered FRB\,191108, this is larger than expected 
from the Milky Way and the IGM \citep{masui-2015b}. The authors argued 
that its high RM and scattering properties suggested a dense 
magnetised environment local to the source. The FRB with the highest published DM, FRB\,160102, had an RM of $-220$\,rad\,m$^{-2}$ \citep{caleb2018b}; its local RM could be as large as $-2400$\,rad\,m$^{-2}$ if a significant portion of the DM comes from the IGM. 
During Breakthrough Listen observations on the Parkes telescope, 
FRB\,180301 was detected and full-polarisation data was preserved \citep{price2019}. They report an RM of 
$-3163\pm20$\,rad\,m$^{-2}$, although the patchiness of their frequency spectrum 
causes the authors to question their Faraday rotation fit. 
CHIME has found a repeating FRB whose RM exceeds the Galactic 
foreground by two orders of magnitude, with RM=$-499.8\pm0.7$\,rad\,m$^{-2}$ \citep{chime2020-repeaters}. Finally, 
FRB\,121102 has an RM of $\sim$\,$10^5$\,rad\,m$^{-2}$ 
and is spatially coincident with a 
bright, compact radio source \citep{michilli2018}. This is larger 
than even the Galactic center magnetar, PSR J1745$-$2900, 
with RM\,$\sim$\,$7\times10^4$\,rad\,m$^{-2}$ \citep{eatough2013}. Both FRB\,121102 and PSR J1745$-$2900 have been seen to exhibit significant
RM variation over month to year timescales \citep{desvignes2018}. 

The analogy between FRB\,121102 and the Galactic 
center magnetar may extend beyond just phenomenological 
similarities. If the persistent radio source coincident 
with FRB\,121102 is similar to a low-luminosity
active galactic nucleus (LLAGN), 
then that system may be another example 
of a circumnuclear magnetar, a scenario  
that has been proposed as a progenitor 
theory of FRBs \citep{pen15}. Alternatively, the radio nebula could correspond to a supernova remnant, magnetar wind nebula, or HII region. Such local environments have been invoked  
as a way to provide local RM, DM, and scattering 
\citep{connor-2016a, piro2016, murase16, piro-2018, margalit18, straal-2020}. 
In each of these cases, it is difficult to predict the distribution 
of observed RMs, but it is likely that the distribution would be broad.
For example, in the circumnuclear magnetar model, the FRB RM 
is a strong function of its distance from the massive black hole.
In young magnetar or supernova remnant models, the RM is expected to change 
with time, and the value depends on when in the  progenitor life cycle  
the FRB was observed. Thus, moderately large RMs like those of FRB\,191108, FRB\,110523 \citep{masui-2015b}, and FRB\,160102 \citep{caleb2018b} may come from a 
similar environment to \mbox{FRB\,121102}.

\subsection{Repetition constraints}
\label{sect-repetition}
Given the extreme local environment of FRB\,121102 and its anomalously high 
repetition rate, it may be asked if frequent repeaters are more 
likely to live near dense magnetised plasma. CHIME recently discovered a 
repeating FRB whose RM is $-499.8\pm0.7$\,rad\,m$^{-2}$, which is roughly 
two orders of magnitude larger than the expected Milky Way contribution in that direction \citep{chime2020-repeaters}. But \citet{chime2020-repeaters}
also report a repeater with RM=$-20\pm1$\,rad\,m$^{-2}$, and most of 
the RM=$-114.6\pm0.6$\,rad\,m$^{-2}$ from another CHIME repeating source, FRB\,180916.J0158+65, is thought 
to be from the Milky Way \citep{chime19-8repeaters}.

We observed the field of \mbox{FRB\,191108} for 120\,hours 
between July 2019 and December 2019 with Apertif, but had 
no repeat detections. 
Apertif has detected and studied other repeating FRBs \citep{2019arXiv191212217O}.
Assuming repetition statistics described by a homogeneous Poisson process, 
our non-detection provides  a $3\,\sigma$ upper-limit on the repeat rate of $3\times10^{-2}$
per hour. We caution, however, that the assumption of stationarity is known to not be valid for some 
FRBs, which show time-variability in their repetition rate \citep{spitler2016, oppermann-2018, gourdji2019}
thereby increasing the probability of seeing zero repeat bursts during follow up \citep{connor-2016b}.

We plan to continue follow-up efforts on the same field, 
which we can do commensally with our full-FoV blind FRB search. The source 
is currently localised to an ellipse with semi-minor and 
semi-major axes of $2.5\,\arcsec$ and $3.5\,\arcmin$, respectively, as described in 
Sect.~\ref{sect-localisation}. If we 
detect \mbox{FRB\,191108} again at a different hour angle than the initial detection, 
we will have several arcsecond localisation in both directions, because the TABs rotate 
as a function of parallactic angle. 


\section{Conclusions}\label{sect-conclusions}
We have reported the detection of a bright, highly Faraday rotated 
FRB in the direction of Local Group galaxy M33 using Apertif. 
By combining multibeam and interferometric information
we were able to localise \mbox{FRB\,191108} 
to a narrow ellipse with radii of 2.5$\,\arcsec$ and 3.5$\,\arcmin$.
The impact parameter with M33 is just 18\,kpc, roughly the 
diameter of that galaxy's disk.
The shared plasma in the halos of M33 and M31 likely contributed to the DM of the FRB,
but not to its scattering, Faraday rotation, or scintillation.
Still, the RM of +474$\pm3$\,rad\,m$^{-2}$ is one of the largest of any 
published value and is an order of magnitude larger 
than the expected contribution from the Milky Way, 
the IGM, and these halos.
The most plausible location of the magnetised plasma is therefore a dense region near the FRB-emitting source itself.


\section*{Acknowledgements}
We thank Dan Stinebring, Jason Hessels, and Monica Relano Pastor
for helpful conversations.
This research was supported by 
the European Research Council under the European Union's Seventh Framework Programme
(FP/2007-2013)/ERC Grant Agreement No. 617199 (`ALERT'), 
and by Vici research programme `ARGO' with project number
639.043.815, financed by the Dutch Research Council (NWO). 
Instrumentation development was supported 
by NWO (grant 614.061.613 `ARTS') and the  
Netherlands Research School for Astronomy (`NOVA4-ARTS' and `NOVA-NW3').
EP acknowledges funding from an NWO Veni Fellowship.
AS and DV acknowledge support from the Netherlands eScience Center (NLeSC) under grant ASDI.15.406.
SMS acknowledges support from the Netherlands Research School for Astronomy (NOVA4-ARTS), and was supported by the
National Aeronautics and Space Administration (NASA) under grant number NNX17AL74G issued through the NNH16ZDA001N
Astrophysics Data Analysis Program (ADAP).
EAKA is supported by the WISE research programme, which is financed by NWO.







\bibliography{ref}
\bibliographystyle{mnras}





\label{lastpage}
\end{document}